\documentclass[conference]{IEEEtran}
\IEEEoverridecommandlockouts

\usepackage{cite}
\usepackage{amsmath,amssymb,amsfonts}
\usepackage{algorithmic}
\usepackage{graphicx}
\usepackage{textcomp}
\usepackage{xcolor}
\def\BibTeX{{\rm B\kern-.05em{\sc i\kern-.025em b}\kern-.08em
    T\kern-.1667em\lower.7ex\hbox{E}\kern-.125emX}}

\usepackage{dsfont}
\usepackage{url}
\usepackage{adjustbox}
\usepackage{graphicx}
\usepackage{ifthen}
\newboolean{showcomments}
\setboolean{showcomments}{true}

\newboolean{showplots}
\setboolean{showplots}{false}

\ifthenelse{\boolean{showcomments}}
{ \newcommand{\mynote}[3]{
   \fbox{\bfseries\sffamily\scriptsize#1}
   {\small$\blacktriangleright$\textsf{\emph{\color{#3}{#2}}}$\blacktriangleleft$}}}
{ \newcommand{\mynote}[3]{}}

\begin{document}

\makeatletter
    \newcommand{\linebreakand}{%
      \end{@IEEEauthorhalign}
      \hfill\mbox{}\par
      \mbox{}\hfill\begin{@IEEEauthorhalign}
    }
    \makeatother

\title{TIMBRE: Efficient Job Recommendation On Heterogeneous Graphs For Professional Recruiters}

\author{
    \IEEEauthorblockN{Éric Behar}
    \IEEEauthorblockA{
        \textit{SAMOVAR, Télécom SudParis}\\
        \textit{Institut Polytechnique de Paris}\\
        91120 Palaiseau, France \\
        eric.behar@telecom-sudparis.eu}
    \and
    \IEEEauthorblockN{Julien Romero}
    \IEEEauthorblockA{
        \textit{SAMOVAR, Télécom SudParis}\\
        \textit{Institut Polytechnique de Paris}\\
        91120 Palaiseau, France \\
        julien.romero@telecom-sudparis.eu}
    \and
    \IEEEauthorblockN{Amel Bouzeghoub}
    \IEEEauthorblockA{
        \textit{SAMOVAR, Télécom SudParis}\\
        \textit{Institut Polytechnique de Paris}\\
        91120 Palaiseau, France \\
        amel.bouzeghoub@telecom-sudparis.eu}
    \linebreakand
    \IEEEauthorblockN{Katarzyna Wegrzyn-Wolska}
    \IEEEauthorblockA{
        \textit{EFREI, 75000 Paris, France} \\
        katarzyna.wegrzyn@efrei.fr}
}

\maketitle

\begin{abstract}
 
Job recommendation gathers many challenges well-known in recommender systems. First, it suffers from the cold start problem, with the user (the candidate) and the item (the job) having a very limited lifespan. It makes the learning of good user and item representations hard. Second, the temporal aspect is crucial: We cannot recommend an item in the future or too much in the past. Therefore, using solely collaborative filtering barely works. Finally, it is essential to integrate information about the users and the items, as we cannot rely only on previous interactions. This paper proposes a temporal graph-based method for job recommendation: \textbf{TIMBRE} (Temporal Integrated Model for Better REcommendations). TIMBRE integrates user and item information into a heterogeneous graph. This graph is adapted to allow efficient temporal recommendation and evaluation, which is later done using a graph neural network. Finally, we evaluate our approach with recommender system metrics, rarely computed on graph-based recommender systems.
\end{abstract}


\section{Introduction}

Today's job market is extremely dynamic and competitive, particularly in the IT sector. One consequence is the multiplication of applicants for each position opening~\cite{talentworks}, leading to a heavy workload for companies. Therefore, they often decide to externalize the process to specialized recruiting firms that can handle many candidates and a wide range of skills in the market. Still, even in these firms, the recruiters have to filter many applicants, leading to two behaviours. First, they resort to automatic ATS (Applicant Tracking System) software that parses the resumes and filters the candidates using simple rules like keyword matching. Second, they often prefer to reverse the process by directly head-hunting good candidates and gathering relevant information about them. Then, when they receive a new job opening, they first look in their database.

Many works propose building recommender systems to assist with the recruiting process~\cite{qin2020recruitment, roy2020recruitment, freire2021recruitment, giabelli2021skills2jobrecruitment, yang2022recruitment, wu2023recruitment}. However, most of them use direct applications from the candidates as training data, leading to noisy input. Very few works were trained on real-life data annotated by professional recruiters. Yet, this kind of recruiter-oriented recommender system can greatly impact the productivity of recruiting firms. Building recommender systems for job openings encounters many challenges that are also present in other kinds of recommendations but are often emphasized in this case. First, the \textbf{cold start problem} is recurring in almost all recommendations and is the focal point of many work on recommender systems in the literature \cite{dong2020mamocoldstart, lu2020metacoldstart, zhu2021learningcoldstart, wei2021contrastivecoldstart, cai2023usercoldstart}. The candidates and the job openings have a \textbf{short lifespan} of a few weeks. Therefore, we have few training points and cannot rely on previous interactions. On the contrary, for more mainstream recommendations like movie recommendations, we generally assume that both the user and the item are here to stay for quite some time.

\begin{figure*}
    \centering
    \includegraphics[width=\textwidth]{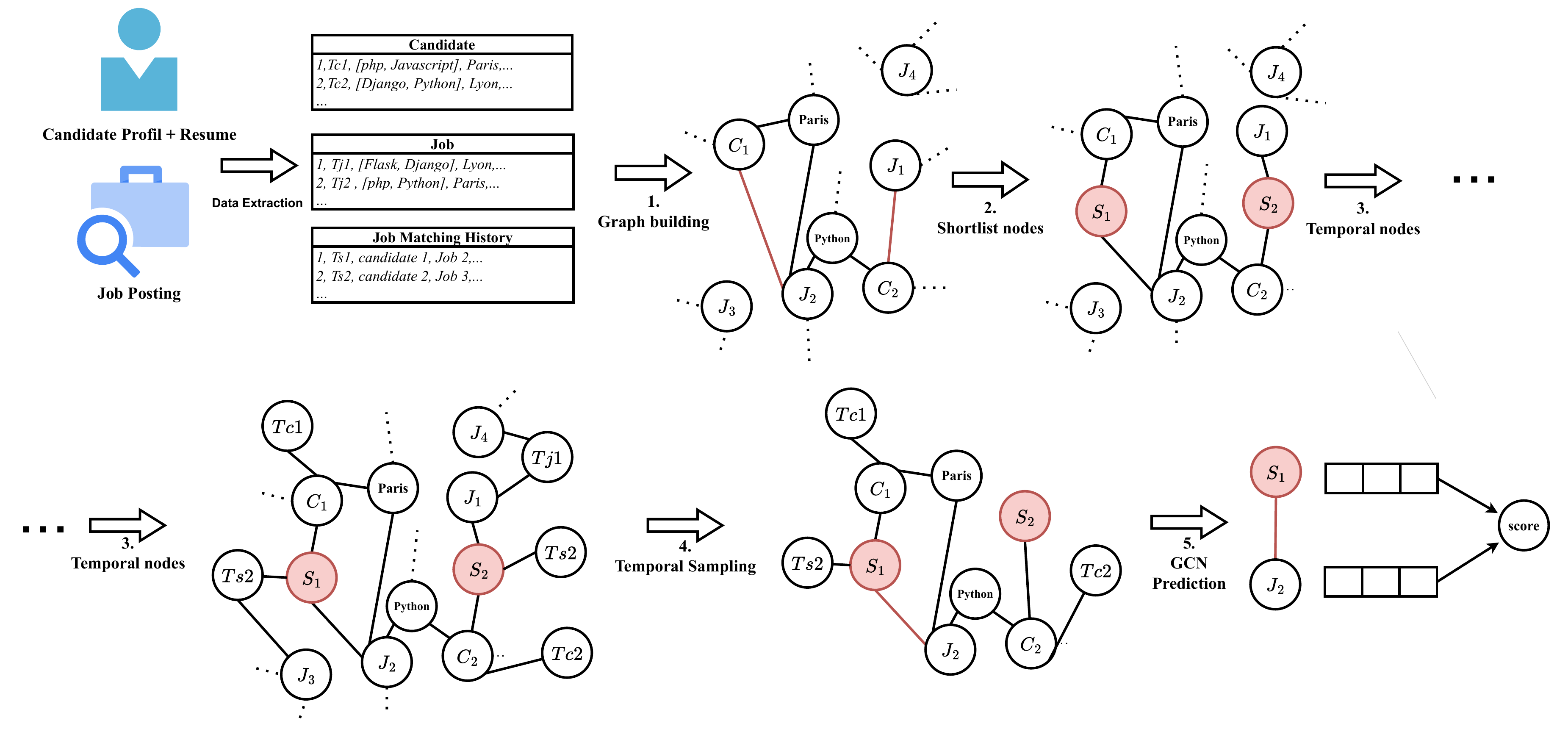}
    \caption{Our complete job recommender system pipeline. 1. We turn our input data into a heterogeneous graph. 2. We replace relations between candidates and jobs with a shortlist node. 3. We add temporal nodes. 4; We apply our temporal sampling algorithm. 5. We apply a graph convolution network and make a prediction based on the representations of the shortlist node and the job node.}
    \label{fig:pipeline}
    \vspace{-0.5cm}
\end{figure*}

Second, because of this limited lifespan, the \textbf{temporal dimension is crucial}. Given a candidate (or a job), we cannot recommend an item in the future (mainly a problem during training) or too far in the past (as the position is likely to be already filled). Therefore, we need to adapt our representations to embrace the dynamicity of the data and not bias the results, particularly during training. The temporal component was often studied in the literature \cite{trivedi2018dyrep, rossi2020tgn, Xu2020Inductive, xu2021tgan}, but often with an assumption that does not hold in our case: User preferences evolve through time. This assumption makes sense when recommending a book or a movie but not for job applications because of the limited lifespan of the user and the item. Therefore, there is \textbf{no need to model the temporal evolution of user and item} representations, as it will only lead to more noise. Finally, as we cannot rely enough on previous interactions, we must strongly emphasize \textbf{additional information} about the users and the items. This means that we need to extract information from various sources and structure them in a way that is exploitable by a recommender system.

To solve these challenges, we introduce \textbf{\textit{TIMBRE}} (Temporal Integrated Model for Better REcommendations), a temporal graph-based recommender system. TIMBRE first ingests data from multiple sources (resumes, job descriptions, recruiter notes, external knowledge bases) and structures them into a \textbf{unified heterogeneous graph}. Then, it adapts this graph to facilitate temporal recommendation. Next, it runs a graph neural network (GNN) to generate a score for a user-item pair. The particularity of this GNN is that it emphasizes the graph's structural aspect rather than the node's representation to counter the effect of the cold start. The GNN is trained by considering the temporal dimension to avoid training bias (mainly from the recruiters~\cite{behar2023}).

In the end, we propose to evaluate our approach using traditional recommender system metrics. Although this should be automatic, most of the literature on graph-based recommendations ignores them as they are difficult to compute. Instead, they prefer metrics based on negative sampling, which are far from reliable. We implemented our evaluation on several baselines and compared them with TIMBRE, showing a clear advantage for our approach in the case of job recommendation.

To summarize, our contributions are the following:
\begin{enumerate}
    \item Extraction and representation of information into a unified temporal heterogeneous graph.
    \item Graph adaptation for a temporal recommendation through reifying the temporal interaction and introducing temporal nodes.
    \item Time-dependent training and evaluation of our GNN using a sampling method boosting collaborative filtering.
    \item Evaluation of a graph-based solution with recommender system metrics.
\end{enumerate}

\section{Previous Work}

\paragraph{Job Recommendation} Many works in the literature tackle job recommendation as a user-item recommendation scenario and thus use a collaborative filtering approach where a user gets recommended the jobs of similar users~\cite{yang2017collaborative,Mishra2020collaborative, dhameliya2019collaborative, prince2023collaborative}. Even though more and more data are produced, they remain primarily inaccessible to private research due to ethical concerns, privacy laws, and strategic concerns. Some effort has been made to provide an anonymized dataset, such as the Xing dataset \cite{recsys2016}, but it is now unavailable. Some works~\cite{covington2016deep,barkan2016item2vec} include additional features during the recommendation; however, they often require a lot of engineering. Some approaches emphasize helping the human recruiter review the candidate's profile by performing resume screening using natural language processing~\cite{daryani2020resumescreening, sinha2021resumescreening, bharadwaj2022resumescreening} or develop resume parsing framework~\cite{sajid2022parsing, mohanty2023parsing, tallapragada2023parsing}. These methods have two limitations. First, they only work if you have a specific job with a pre-selected pool of candidates. This does not work for proactive job recommendations such as headhunting. Secondly, it implies that the information in the resume is accurate. In our context, candidates' profiles are composed of a resume and information collected by an expert recruiter during an interview with the candidate. We must also mention some recent work shows promising results to either improve part of a recommender system or let a large language model (LLM) decide on job-candidate match~\cite{li2023llm, ghosh2023jobrecogpt, wu2024llm}. The results are, however, limited to small-scale scenarios, as properly ingesting thousands of resumes is a challenging task for the current LLM frameworks.

\paragraph{Graph-Based Recommendation} Graph-based approaches to recommendations are equivalent to the link prediction task: Given a graph, we want to predict whether there will be a connection between a user and an item. The advantage of homogeneous and heterogeneous graphs is that they can represent connected data and semantic information that can be used to make recommendations~\cite{wu2022graph,guosurvey2020, wang2021graph,WuGNNSurvey2023} using graph neural networks (GNN), even in the case of jobs~\cite{shalaby2017help}. However, the construction of the graph can be problematic due to the absence or few numbers of features~\cite{WangKGAttention2019,YANGHieAttentionGCN2020}, which sometimes leads to using an external knowledge base like DBpedia for famous entities~\cite{palumbo2017recsys}. Concerning the architecture, we find variations on top of graph convolutional networks~\cite{guosurvey2020,WuGNNSurvey2023,behar2023} and graph attention networks~\cite{WangKGAttention2019,YANGHieAttentionGCN2020}.

\paragraph{Temporal Recommendation} In many applications, it is crucial to model the change in user preferences for long-term and short-term modifications~\cite{rich1983users,bogina2023considering}. In the literature, several techniques are used to model the user representation change through time. We can cite those using latent Dirichlet allocation
~\cite{kowald2015refining}, deep learning techniques~\cite{song2016multi}, reinforcement learning~\cite{wang2014exploration}, matrix 
factorization~\cite{white2010modeling}, recurrent neural networks (RNN)~\cite{yu2016dynamic}, or Markov chains~\cite{he2016vista}. We find similar techniques for temporal graph recommendations. The Neighborhood-Aware Temporal network~\cite{luo2022nat} (NAT) stores the temporal modifications of the neighbor of a node in a dictionary and then uses an RNN to make the representation of a node evolve. Temporal Graph Network (TGNs)~\cite{rossi2020tgn} makes the embeddings of each node evolve through time using a graph attention network and an indication of the time delta since the previous interaction. A variant of the temporal recommendation is the sequential recommendation, where the goal is to predict the next interaction~\cite{wu2020sse,sun2019bert4rec,kang2018self}. However, this setup mostly disregards the time of the interactions and can only make predictions with enough previous interactions (often at least three), which avoids the problem of the cold start.

\paragraph{Temporal Graph Neural Networks} Many applications consider the temporal dimension in a graph, mainly through the representation of an event stream as a graph. For spatio-temporal 
events, the temporal dimension is either used to adjust a distance function defining the neighbors of a node~\cite{yang2024evgnn,Schaefer_2022_CVPR} and to create graph snapshots that contain all the events in a time window~\cite{bi2020graph}.

\paragraph{Datasets for temporal recommendation with side information}
Some public datasets exist for temporal recommendation~\cite{harper2015movielens,cho2011friendship,asghar2016yelp}. However, due to privacy limitations, they often come with very limited side information, especially for the users. Besides, they often supposed that most users and items have enough recommendations to make relevant recommendations. However, this is not the case for job recommendations, which makes it necessary to develop new techniques that can better balance external information and interactions.
\section{Problem}

\paragraph{Background} In this paper, we will consider heterogeneous graphs. A heterogeneous graph $G$ is defined by a tuple $(V, R, D, E, \Omega)$ where V is a finite set of nodes, $R$ is a finite set of relationships, $D$ is a set of types, $E \subset V \times R \times V$ is the set of edges, and $\Omega \subset V \times D$ is the set of types associated to a node. We can associate properties with the nodes or edges of a heterogeneous graph using a function $P(v_1)$ or $P(v_1, r, v_2)$ where $v_1 \in V$, $v_2 \in V$, and $r \in R$. In this paper, we only consider properties for nodes. We talk about a \textbf{temporal heterogeneous graph} when the property represents temporal information. This information generally represents the date of a node or edge creation.

\paragraph{Our Problem} This paper uses a dataset of users (candidates) $U$ and items (job postings) $I$. We suppose we can access a textual document for each user and item. In practice, the document will be a resume, information from recruiters for the candidates, and a job description for a job posting. Then, the dataset contains a set of $N$ interactions $(u, i, t) \in U \times I \times T$ where $T$ represents the timestamps of the interactions (e.g., a POSIX time). In the real world, an interaction means that a user $u$ was selected for a job $i$ by a recruiter at time $t$. We call the selection of a candidate \textbf{shortlisting}.

The problem we tackle is the following: Given a user $u \in U$ and a time $t \in T$, we rank all the items $i \in I$ such that the higher the rank, the better the recommendation at time $t$.

In our case, as we will explain in Section~\ref{sec:graph-construction}, we represent a user $u \in U$ and a time $t \in T$ by a new entity called a shortlist $s$. Our problem becomes to rank all the items $i \in I$ for a given shortlist $s$. We kept we original problem for the baselines without this shortlist entity.

This problem slightly differs from previous works in several aspects. First, it is widespread to focus on predicting the next interaction (sequential recommendation), ignoring the current time. However, the formulation of our problem makes the training phase easier, as we will see later. Besides, in practice, we are concerned about recommending when the recommendation is required. In our case, a job has a limited lifespan. Second, many works on temporal recommendation on graphs only focus on classifying a random negative sample and a true example. As we will notice later, this evaluation's results are unsuitable for recommender systems.
\section{Methodology}

Figure~\ref{fig:pipeline} gives an overview of TIMBRE.

\subsection{Graph Construction}
\label{sec:graph-construction}

\paragraph{Basic Structure} As input to our algorithm consists of candidate resumes, job postings, and information prefilled by the candidate or the recruiter. After discussing with professional recruiters, we selected features and represented them as a heterogeneous temporal graph (similar to~\cite{behar2023}). More specifically, we have \textbf{eleven kinds of nodes}: candidates, jobs, companies, salaries, number of years of experience, skills, skill concepts (high-level skills), types of contract (permanent, temporary, freelance), location (through a zip code), job category, and candidate origins (recruiting platform, like Linkedin). All these fields are completed manually as part of the recruitment process of a company (either by the candidate when they apply for a position or by a recruiter when they enter a new position or candidate in the database), except for the skills, which are augmented automatically by searching for keywords in the resumes and job descriptions. These keywords come from two \textbf{external knowledge bases}: the European classification of Skills, Competencies, Qualifications, and Occupations \cite{de2015esco} and Wikidata \cite{Vrande2014wikidata}. They also enrich the information about the skills by introducing a hierarchy of skills. The nodes of our graph are connected with named relations like \textit{hasSkill} or \textit{atLocation}. For nodes with temporal information (date of creation for the candidates and the jobs), we encode it as a timestamp in a property of the node. We set the timestamp to 0 for others to make temporal sampling easier. We also have a resume and a job posting as text, so we turned them into embeddings using a sentence transformer model~\cite{wang2020minilm} and attached them to the nodes. In the end, our graph contains \textbf{44 different edge types} (half are, in fact, reverse edge types).

\paragraph{Shortlist Nodes} Most works represent a recommendation between a user and an item by an edge. However, it might cause problems for a temporal recommendation. First, \textbf{encoding the interaction time directly on an edge is hard}, and we cannot use the candidate and job nodes as they already have timestamps. Second, \textbf{generating a negative sample becomes harder}. What is the timestamp of the negative interaction, and where do we encode it, as the fake interaction is not part of the graph? To solve these problems, we created a new kind of node inspired by reification principles used in RDF (Resource Description Framework). For each interaction, $(u, i, t)$, we create a new node $S_{u,i,t}$ called a \textbf{shortlist node} that is connected to $u$ and $i$, has $t$ as a timestamp and is linked to the corresponding temporal node. We do not have a direct connection between $u$ and $i$. Now, if we want to create a negative sample for an interaction $(u, i, t)$, we pick a random item $i'$ and connect it to $S_{u,i,t}$. \textbf{The time of the negative sample is automatically managed}. Our final graph has \textbf{no edge between a candidate and a job}. The interaction goes through a shortlist node.

\paragraph{Temporal Nodes} In most works, the temporal information is encoded by the nodes' embeddings depending on time. This article chose a simpler yet effective approach. The timestamp property of the nodes will be used for sampling (see later). Besides, time is crucial when recommending a job, as the job might be too old. Therefore, we introduced a \textbf{new type of node representing the number of months since the first shortlist}. This node has a feature composed of a single number, the number of months, and a timestamp corresponding to the time at the beginning of the month. \textbf{Temporal nodes} are connected to candidates, jobs, and shortlist nodes.

\subsection{Job Prediction}

We train our network using the \textbf{link prediction} task. The goal of this task is to predict whether there is a link or not between two nodes in the graph. In our case, we want to predict whether there is a \textbf{link between a shortlist node and a job node}, equivalent to recommending a job for a candidate at a given timestamp. To recommend a candidate for a job, we can predict a link between a shortlist node and a candidate node. For the training part, we need to have positive samples (coming from the dataset) and negative samples (generated randomly as explained in Section~\ref{sec:graph-construction}).

\begin{figure}
    \centering
    \includegraphics[width=0.5\linewidth]{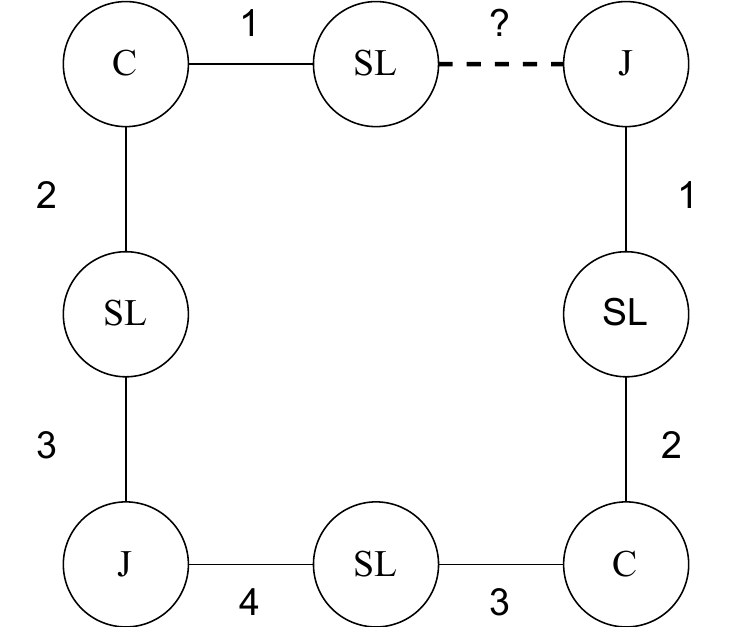}
    \caption{Sampling Distance For CF. C=Candidate, J=Job, SL=Shortlist.}
    \label{fig:sampling-distance}
    \vspace{-0.5cm}
\end{figure}

Because of the size of the graph, using the entire graph to make the recommendation is too expensive. Therefore, following previous works~\cite{Xu2020Inductive}, we decided to \textbf{sample a sub-graph} around the shortlist node and the job node we consider. During this sampling, it is crucial to ignore the nodes that do not exist at the moment of the recommendation, i.e., we can only keep the candidates, job postings, and shortlisting events anterior to the current shortlist node we consider. This kind of filtering is not necessarily done in the literature, and it causes problems in the case of data annotated by recruiters as they tend to create groups of candidates and submit them to the same jobs~\cite{behar2023}. Therefore, the component of our sampling strategy includes a \textbf{temporal filtering of future events}.

In most cases, sampling a sub-graph containing the nodes at a distance two or less from the original nodes gives good results. However, in our case, given the diversity of edge types and the presence of the shortlist node, we need to pay close attention. Particularly, if we want to have a form of collaborative filtering (CF), we need to sample candidates who applied for similar jobs. In our case, we need to have a sampling depth of four (see Figure~\ref{fig:sampling-distance}), but we need to be careful not to sample certain edge types. Indeed, if we follow the edges going from a skill to a candidate, we will sample too many nodes (all the candidates with that skill). Therefore, we used a \textbf{selective sampling} by only sampling edges going from a candidate, shortlist, or job node to another node (23 edge types over 44 in total. E.g., the edges (shortlist, has\_application, job), (candidate, has\_skill, skill), or (job, has\_experience, experience)). Besides, we decided to \textbf{take all the edges at a certain depth} and not a sample. Although it makes the computation longer, it introduces less noise in the results and makes the sampling deterministic.

\subsection{Graph Architecture}

This paper uses a graph neural network (GNN), more precisely a \textbf{graph convolutional network} (GCN), to make the predictions. Each node in our graph is associated with an embedding that does not depend on time. For nodes with features (candidates, jobs, time nodes), the final embedding is a linear combination of a learned embedding and the feature vector. Then, we have several layers of SAGE convolutions~\cite{Xu2020Inductive}, where each of them is normalized using a layer normalization~\cite{ba2016layer}. The non-linearity function is a GELU (Gaussian Error Linear Unit)~\cite{hendrycks2016gaussian}. After the convolutions, we get a vector for the shortlist node and the job we consider, and we compare them using cosine similarity. Finally, we apply the binary cross entropy loss. As we have a heterogeneous graph, we used the transformation from~\cite{schlichtkrull2018modeling} to adapt our network.

\subsection{Evaluation}

Most graph-based approaches from the literature~\cite{rossi2020tgn,luo2022nat,kumar2019jodie,xu2021tgan} only report metrics like \textbf{precision, recall, and area under the curve (AUC)} on the task of link prediction. This biased evaluation gives minimal insight into the model's performance. Indeed, these metrics \textbf{only evaluate the capability to separate negative samples from positive samples}. The negative samples are often drawn randomly, and these random items are effortless to differentiate from positive examples. Therefore, the metrics reported in previous works are very high but without much interest.

Instead, \textbf{we evaluate the capability of the models to rank the items for a given user}. With a matrix-based approach with fixed precomputed embeddings of each user and item, the ranking is easy and fast to compute: We perform a matrix multiplication and sort the results. However, this is impossible for graph-based approaches trained for link prediction. Instead, we must \textbf{run the link prediction task for each user-item pair} many times and sort the results. This process can be very long compared to the training, explaining partially why the literature abandoned the ranking evaluation. Another reason is that, \textbf{for temporal graphs}, when predicting a user-item interaction, we need to decide when this interaction happens, and we go back to the problem raised in Section~\ref{sec:graph-construction}. We need to use the timestamp of real interactions, but it is unclear what to pick for false interactions. \textbf{Using our shortlist node solves the problem}. We focus on predicting a user-item interaction at a given time that is directly encoded in the node. Therefore, we do not have to care about the time, and we do as if the task was to rank all the jobs for a shortlist node.

\section{Experiment Setup}

\paragraph{Dataset} We used the JTH (Job Tracking History) dataset introduced in~\cite{behar2023}. This dataset comprises 67k real candidate-job associations manually annotated by professional recruiters. We can access 67k candidates (only 26k have at least an associated job) and 4k jobs (most have a recommendation). Candidates have a resume, and jobs have a description. Besides, recruiters might add additional information like relevant skills or wanted salaries. We divided our dataset into train, validation, and test sets \textbf{using a temporal order} to prevent data leakage, with a proportion of 80/10/10. Due to the nature of the data, the testing dataset mostly contains \textbf{unseen users}, making the cold start problem central.

\paragraph{Baselines}

We compared our approach with the following state-of-the-art approaches: Temporal Graph Network~\cite{rossi2020tgn} (TGN), Neighbour Aware Temporal Network~\cite{luo2022nat} (NAT), JODIE~\cite{kumar2019jodie}, DYREP~\cite{trivedi2018dyrep}, TGSRec~\cite{fan2021continuous}. Besides, we have two approaches based on large language models (LLM) that compute an embedding for each user and item and then rank the item using the cosine similarity in a deterministic way (therefore, no standard deviation). We used text-embedding-3 from OpenAI~\cite{textEmbeddings} and BGE-M3~\cite{bge-m3}.

\paragraph{Evaluation Setup} We focused on predicting a job for a given candidate, but the opposite would work the same. Note that for our approach, we actually want to predict a job for a shortlist node, i.e., a job for a candidate at a given time, making our problem harder than the one for the baselines. For simplicity, in what follows, we call ``user'' a normal user for the baselines but a shortlist node in our approach. During the training, all the baselines have access to the training and validation sets. Then, for the final evaluation of the test set, we proceed as follows. For each interaction between a user $u$ and an item $i$, we first start by drawing a random negative sample $i_{neg}$, and we compute the scores for both $i$ and $i_{neg}$. That allows us to get the \textbf{classification metrics} (see below). Then, for each item $i'$ (not necessary in the test set), we compute the score between $u$ and $i'$ to \textbf{produce a ranking of all the items} for the user $u$. This ranking is used to compute the recommendation metrics (see below). Note that for each interaction, the model can have access to all previous interactions but not to future interactions.

\paragraph{Metrics} We will report two kinds of metrics: Metrics related to the classification task with negative samples (as used in~\cite{rossi2020tgn,luo2022nat,kumar2019jodie,xu2021tgan}) and metrics traditionally used for recommendation. We use the area-under-the-curve (\textbf{AUC}) and the \textbf{precision} for the classification metrics. Then, we use the mean reciprocal rank (\textbf{MRR}) and \textbf{Recall@10} for the recommendation metrics. For most experiments, we ran them over 10 different seeds and reported the mean score and the standard deviation (SD). In details, we have:

\begin{equation}
MRR = \frac{1}{|D|} \sum_{i=1}^{|D|} \frac{1}{rank_i}
\end{equation}
\begin{equation}
Recall@K = \frac{1}{|D|} \sum_{i=1}^{|D|} \mathds{1}(rank_i \leq K)
\end{equation}

$D$ is the test dataset, and $rank_i$ is the rank of the first positive answer. Note that these formulas work only in our case, as we have exactly one positive example by shortlist node (by construction). The formulas were adapted for the baselines, which do not use the shortlist node.


\paragraph{Configuration} We wrote our code in Python, using Torch and Pytorch-Geometric~\cite{fey2019fast}. We ran our experiments on an NVIDIA Tesla V100 GPU. A training and evaluation cycle took between one day and two days to run (most of it is the evaluation). The optimizer is Adam~\cite{kingma2014adam} with a learning rate of $1\mathrm{e}{-5}$ and a weight decay of $1\mathrm{e}{-4}$. Our GCN has three layers. Three is a tradeoff between computation time and performance, as adding more layers makes the experiments longer.
We make our code available on \url{https://github.com/Aunsiels/job_recommendation}. For the baselines, we reused the code provided by the original authors and adapted our data to fit their input format. Besides, we created a feature vector for each candidate and job to help the baselines.
\section{Results}

\paragraph{Main Results} \ifthenelse{\boolean{showplots}}{Figures~\ref{fig:sub1-main-results} and~\ref{fig:sub2-main-results}}{Table~\ref{tab:main-results}} displays the comparison of our approach (TIMBRE) with the different baselines. We observe significant variations when looking at the metrics reported originally with the baselines (AUC and precision). Two methods seem to have an edge: Jodie and TIMBRE. However, \textbf{we do not observe the same behavior when looking at the recommendation metrics} (MRR and Recall@10). TIMBRE significantly outperforms all the other baselines, with a factor of 5 for the MRR and Recall@10. Besides, looking solely at the baselines that compute the AUC and the precision, TGSRec got the best score, which was not true for the AUC and precision. Again, it shows that these non-recommendation metrics are helpful during training but do not fully indicate the system's final performance. Finally, we must note that the scores underestimate the model's performance in production. In practice, we can filter many results using simple filters, like checking if a position is still open. This also makes the generation of the recommendation faster. However, we wanted to test the capability to understand and model business rules and temporal events. Besides, we did not necessarily have access to all the data to write these filters.

The two main reasons for our approach's success are its capability to \textbf{integrate external information in a unified system} and its possibility to \textbf{leverage recent interactions without retraining} (not feasible in practice due to computation time), even in the test case. Therefore, we have much more success in tackling the cold start problem, which is generalized in the case of job recommendation, as discussed earlier.

\ifthenelse{\boolean{showplots}}
{

\begin{figure}
    \centering
    \includegraphics[width=\linewidth]{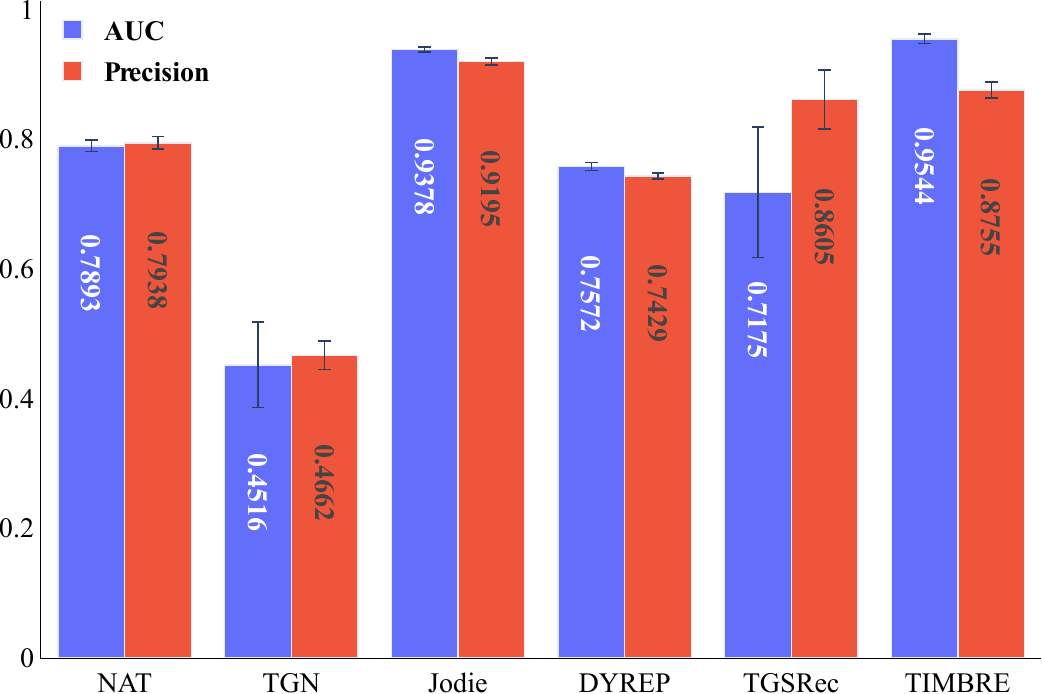}
    \caption{Classification metrics computed on the negative sample}
    \label{fig:sub1-main-results}
\end{figure}

\begin{figure}
    \centering
    \includegraphics[width=\linewidth]{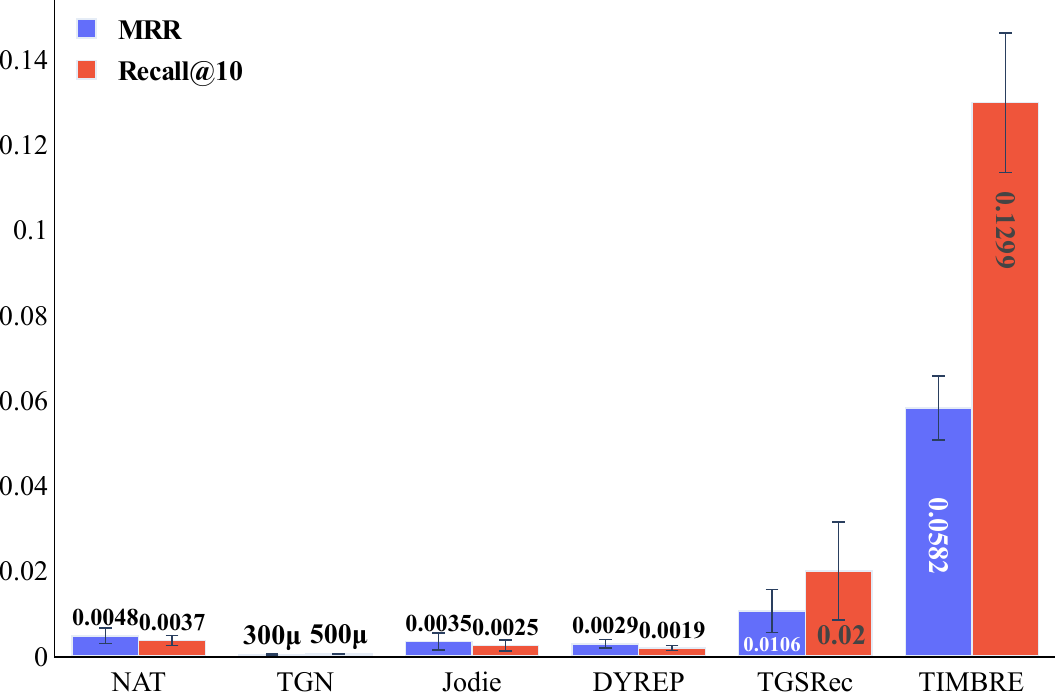}
    \caption{Recommendation system metrics}
    \label{fig:sub2-main-results}
\end{figure}

}
{ \begin{table*}[ht]
    \centering
    \adjustbox{max width=\textwidth}{
    \begin{tabular}{|l|c|c|c|c|c|c|c|c|}
\hline
Metrics & NAT & TGN & Jodie & DYREP & TGSRec & OpenAI & BGE-M3 & TIMBRE \\
\hline
AUC & 0.7893 (SD 0.0091) & 0.4283 (SD 0.0607)& 0.9356 (SD 0.0041)& 0.7489 (SD 0.0052)& 0.7175 (SD 0.1007) & - & - & \textbf{0.9479} (SD 0.0089)\\
Precision & 0.7938 (SD 0.0097) & 0.4579 (SD 0.0116)& \textbf{0.9142} (SD 0.0008)& 0.7375 (SD 0.0046)& 0.8605 (SD 0.0457) & - & - & 0.8640 (SD 0.0112)\\
\hline
MRR & 0.0048 (SD 0.0018) & 0.0004 (SD 0.0001)& 0.0049 (SD 0.0026)& 0.0022 (SD 0.0026)& 0.0106 (SD 0.0051) & 0.0117 (SD 0)& 0.0173 (SD 0)& \textbf{0.0909} (SD 0.0270)\\
Recall@10 & 0.0037 (SD 0.0012) & 0.0005 (SD 0.0000) & 0.0029 (SD 0.0014)& 0.0018 (SD 0.0011)& 0.0200 (SD 0.0115) & 0.0200 (SD 0)& 0.0353 (SD 0)& \textbf{0.1965} (SD 0.0501)\\
\hline

    \end{tabular}
        }

    \caption{Comparison with the baselines. SD = Standard deviation. }
    \label{tab:main-results}
    \vspace{-0.5cm}
\end{table*}

 }

\paragraph{Ablation Study} To understand which components were helpful or not,  we performed an ablation study that we reported in Table~\ref{tab:ablation_study}. The first thing to realize is that a significant component of our system is the \textbf{temporal nodes}. Although very simple (compared to the complex modeling in the baselines), they allow the model to discard jobs that are too old. Next, we observe that only one feature seems useless: The number of years of experience. Looking more closely at the data, we notice that most job postings (96\%) do not have this information filled. Therefore, this feature creates noise. Removing the other feature harmed the results, which was expected. The categories (a manual classification of the domain of expertise made by the recruiter) seem to be the most essential feature. The reason is that it is pretty clean and has limited possible values. The zip code is also crucial, as we mostly want to recruit people near the job posting. The candidate's origin plays a significant role, which indicates that some sources of candidates are more reliable than others. As we could have guessed, the type of contract is also essential to know. However, knowing the recruiting company is not that helpful. Surprisingly, the salary feature is not that central. The reason is similar to the years of experience: Very few candidates choose to provide that information that information. Finally, removing the skills or the concepts seems to have a similar impact. As the concepts represent the hierarchy of skills, we also remove them by removing the skills. So, high-level skills appear more critical when assigning a job than fine-grain skills.

We continued the ablation study by analyzing several scenarios. First, we removed all the features (\nobreakdash-all, including time nodes). We observed that TIMBRE still outperforms several baselines, which shows that it can leverage interactions as well as previous approaches. To understand what we are doing better, we tried to remove the features of the jobs and candidates (\nobreakdash-features) and our collaborative filtering sampling (\nobreakdash-collab.). For this last point, we sampled all the nodes at a depth of two. From the results, we can conclude that removing the features has a slightly negative impact, but \textbf{changing the sampling was the critical point}. It shows that our analysis was correct: We must pick the sampling correctly to ensure we allow collaborative filtering.

\begin{table}[ht]
\adjustbox{max width=\linewidth}{
    \centering
    \begin{tabular}{|c|c|c|c|c|}
    \hline
       \textbf{Setup}  & \textbf{AUC} & \textbf{Precision} & \textbf{MRR} & \textbf{Recall@10} \\
       \hline
       All  & 0.9509& 0.8717	& 0.0763& 0.1651	\\
       \hline
       -experience& 0.9479& 0.8640	& 0.0909	& 0.1965	\\
       -company& 0.9531& 0.8665& 0.0652& 0.1476\\
       -salary& 0.9486 & 	0.8794 & 0.0593 & 0.1258	\\
       -skill  & 0.9470 & 0.8723 & 0.0577 & 0.1245	\\
       -concept& 0.9515 & 0.8551	& 0.0575 & 0.1305 \\
       -contract& 0.9486	& 0.8731	& 0.0547	& 0.1302	\\
       -origin& 0.9424 & 0.8650	& 0.0529	& 0.1133	\\
       -zip&  0.9441	& 	0.8697	& 0.0438	& 0.0945	\\
       -categories& 0.9439	& 0.8731	& 0.0413	& 0.0868	\\
       -temporal nodes & 0.6533 & 0.7005	& 0.0058	& 0.0058	\\
              \hline
       -all & 0.6920 & 0.6678	& 0.0095	& 0.0153	\\
       -all -features & 0.7236& 0.6651& 0.0070 & 0.0118\\
       -all -collab. & 0.7050 & 0.6448& 0.0066& 0.0094\\
       \hline
       -features & 0.9558& 0.8668& 0.0692& 0.1549\\
       -collab. & 0.8877       & 0.7606      & 0.0397      & 0.0745      \\
       \hline
    \end{tabular}
    }
    \caption{Ablation Study}
    \label{tab:ablation_study}
    \vspace{-0.5cm}
\end{table}

\paragraph{Error Analysis} To better understand the results of our experiments, we performed an error analysis. We took a random sample of 100 users in the test set and associated the most probable job position according to our model. Then, we asked a human annotator in a recruiting company to access the recommendations with the candidate profile and job description and manually label them as correct or incorrect. If the recommendation is wrong, the annotator must also give a reason. We also asked the annotator only to use the information in the resume and job description, making the evaluation stricter than what we would expect in real life (we do not account for job progression, for example). Our results are presented in Table~\ref{tab:error-analysis}.
As we can see, the top recommendation is often wrong.
The most frequent cause is a mismatch of skills, which is consistent with Table~\ref{tab:ablation_study} where we saw that removing skills was not that harmful. Interestingly, the job title is often correct (e.g., Frontend developer), but the technologies required do not match. A potential cause of why the skills are not correctly used is how they are extracted. Recruiters rarely fill them, but they are extracted automatically from resumes, thus creating a lot of noise. Besides, the annotator is not necessarily aware of similar libraries or skills that can quickly be acquired in a new position, making the annotation hard. Finally, the graph might not contain enough information to understand a skill. A solution could be to develop a more fine-grained ontology for skills.
Another reason for error is a time inconsistency between a candidate and a position (the candidate was created too much in the past or future). A possible way to solve this problem would be to hardcode a time filtering or to adapt the negative sampling to learn the time constraints better.
Next, we observed issues related to a mismatch in experience: A junior position is often assigned to a senior or team leader person (the opposite is rarely true). As we already mentioned, the experience of a candidate and a job is seldom filled by the recruiter and, therefore, hard to exploit, although it can be found or guessed from the resume.
Finally, we see very few errors due to a wrong location. This field is complex to access as a candidate might be willing to move to a new place. It is also an information that is time-dependent as candidate addresses can become obsolete.
To conclude, all the problems reported here were also present in the baselines, which shows that these recommender systems cannot be used directly out of the box. In a practical case, our system could be included in a broader system with a possible post-filtering to refine the results.

\begin{table}[ht]
    \centering
    \begin{tabular}{|l|c|}
        \hline 
        & Percentage \\ 
        \hline
        Correct                           & 21 \%\\ 
        Incorrect                         &  79 \%\\
        \hline
        Incorrect - Mismatch skills         & 60 \%\\ 
        Incorrect - Incorrect temporality & 14 \%\\ 
        Incorrect - Lack of experience    & 11 \%\\
        Incorrect - Wrong Location        & 4  \%\\
        Incorrect - Overqualified         & 1  \%\\
        \hline
        \end{tabular}
        
    \caption{Error Analysis}
    \label{tab:error-analysis}
    \vspace{-1cm}
\end{table}

\section{Conclusion}

This paper introduces TIMBRE, a temporal job recommender system based on graphs. TIMBRE first integrates all the available information into a temporal heterogeneous graph. Then, it uses three components to facilitate the temporal recommendation and improve performance: The inclusion of a reification node (the shortlist node) that represents an interaction at a given time, the addition of a temporal node that encodes a point in time, and smart sampling that enables collaborative filtering. Our experiments showed that our methodology outperforms state-of-the-art temporal graph recommendation methods, particularly using recommendation metrics that were rarely used before with graphs.

\paragraph{Limitations and Future Works} In this paper, we discuss job recommendations, which are often discrimination-prone. We did our best to remove any feature related to gender or ethnicity. Our solution can also have a societal impact as it automatizes part of the recruitment process. However, we want to stress that it should be used to assist recruiters in finding the best position for a given person, as we do not provide any strong guarantee of the results.

Due to the nature of the data we manipulated, we used a private dataset in our experiment. As there is no equivalent public dataset, applying our approach to another real-life dataset on job recommendation is hard, limiting our evaluation's scope. However, our study gives valuable insights into how recruiting works and how it can be improved and assisted. In particular, although our data is of relatively high quality due to the annotation by professional recruiters, it is still subject to noise from the annotation process, mainly introduced by junior recruiters. In future work, we would like to introduce the full result of the recruiting process (until the contract is signed) to analyze the candidates better and provide ways to improve the recruiting process by giving feedback.

As our ablation study shows, including features in the graph is not necessarily trivial and might require further consideration. For example, although the company is irrelevant, its sector may be interesting. That would allow the inclusion of previous positions occupied by a candidate in the graph. For the sparse features, it might be worth finding a way to fill them. However, it would require significant human labor. A possible future work would be to have a human-in-the-loop system in which we suggest candidates and ask for further information. Concerning the graph sampling strategy, we observed that sampling all the nodes at a given depth (four in our case) and following only certain edge types gave the best results while still keeping reasonable computation times. If execution time is critical or computation resources are limited, we could use more advanced sampling strategies like PASS~\cite{10.1145/3447548.3467284} that sample the most important nodes for our task. It will raise the question of adapting such a sampling for temporal heterogeneous graphs. Finally, we assumed that a candidate's preference does not change, impacting how we model time. Although it is true in most cases, future work could try to include the previous experiences of a candidate to model their career path better.

\bibliography{bib_short}
\bibliographystyle{IEEEtran}

\end{document}